# Coherent manipulation of dipolar coupled spins in an anisotropic environment

E.I. Baibekov[1)*], M.R. Gafurov[1)], D.G. Zverev[1)], I.N. Kurkin[1)], B.Z. Malkin[1)], B. Barbara[2)]

[1)]Kazan Federal University, 420008 Kazan, Russian Federation
[2)]Institut Néel, CNRS, BP166, 38042 Grenoble Cedex 9 and Université Joseph Fourier, France

*Corresponding author. E-mail: edbaibek@gmail.com

**Abstract**

We study coherent dynamics in a system of dipolar coupled spin qubits diluted in solid and subjected to a driving microwave field. In the case of rare earth ions, anisotropic crystal background results in anisotropic $g$ tensor and thus modifies the dipolar coupling. We develop a microscopic theory of spin relaxation in transient regime for the frequently encountered case of axially symmetric crystal field. The calculated decoherence rate is nonlinear in Rabi frequency. We show that the direction of static magnetic field that corresponds to the highest spin $g$-factor is preferable in order to obtain higher number of coherent qubit operations. The results of calculations are in excellent agreement with our experimental data on Rabi oscillations recorded for a series of $CaWO_4$ crystals with different concentrations of $Nd^{3+}$ ions.

**Introduction**

It is well-known that localized electron spins in a solid are potential qubits for quantum information processing [1] since they provide opportunities for scaling and have long coherence times (up to several ms). Among possible implementations are quantum dots [2], NV centers in diamond [3], single-molecule magnets [4-7], and paramagnetic ions diluted in single crystals [8-13]. If the number of paramagnetic particles is large enough ($>10^{12}$), the spin manipulations necessary for quantum computing can be achieved with the standard instrumentation of pulsed electron paramagnetic resonance (EPR) spectroscopy. The crystal sample is placed inside the microwave (mw) cavity of EPR spectrometer. Static magnetic field $B_0$ creates the gap $\omega_0$ between the energy levels of the spin ½. The spin states are controlled using a pulsed mw field $B_1$ of resonant frequency $\omega_0$. Each pulse induces nutations of the spin vector over the Bloch sphere, resulting in oscillating projection of its magnetic moment called Rabi oscillations (ROs [14,15]). If the pulse duration is long enough, a number of oscillations can be recorded. A successful demonstration of long-living ROs is a necessary step before one can implement a given type of spin qubits as a part of a working quantum computer. Note that one should not mix the decay time of the ROs $\tau_R$ (that we further call Rabi time [16]) with the phase memory time $T_2$, since the last one reflects the spin coherence maintained in the absence of the driving mw field.

The ROs that are acquired from the paramagnetic centers diluted in solids decay due to numerous reasons. As follows from our previous research [9,17], the most influential are: (i) dispersion of $\omega_0$ (inhomogeneous broadening of the EPR line), (ii) spatial distribution of $B_1$ in mw resonator, and (iii) magnetic dipole interactions between the paramagnetic centers. The first two result in distribution of nutation frequencies inside the spin ensemble, so that the decay of ROs is caused by the dephasing of the Bloch vectors belonging to different spin packets. In this case, the decay rate is linear in the frequency of ROs $\Omega_R$ (Rabi frequency), which itself is linear in $B_1$. Dipolar interactions, on the one hand, provide entanglement of the states of different spins, which is a vital part of quantum computation process. On the other hand, since these interactions are long-ranged, a given paramagnetic center is coupled simultaneously to a considerable number of other centers in the solid, and the local magnetic field thus produced has random-like character [18]. Because of a reasonable simplicity of the experimental procedure and the ability to control various parameters (intensity of mw field, the spin frequency and concentration, etc.), paramagnetic ions diluted in a solid matrix represent a very convenient system to study decoherence inside the spin ensemble driven by the microwaves.

Until recently, the existing theoretical models accounting for the role of dipolar interactions in the decay of ROs were all based on certain modifications of conventional Bloch equations [18,19], with an attempt to justify the empirical dependence of $\tau_R$ on $\Omega_R$ obtained for $E_1'$ centers in silica and [AlO$_4$]$^0$ centers in quartz [20]:

$$\tau_R^{-1} = \alpha + \beta \Omega_R. \qquad (1)$$

In our recent paper [17], we presented a microscopic model that contained no phenomenological parameters and allowed ab-initio calculation of spin dynamics of dipolar coupled spin ensemble in the microwave-driven regime. It was assumed that the ensemble consisted of the spin particles with isotropic g factor [17-19]. Such assumption is valid if the spins are dispersed in an amorphous medium or in a crystal of cubic symmetry. To the best of our knowledge, no attempt to study theoretically the dependence of $\tau_R$ on the directions of vectors $\boldsymbol{B}_0$ and $\boldsymbol{B}_1$ in the case when the background symmetry is lower than cubic has ever been made. In most cases, the crystal anisotropy does not contribute much to the *g* factor of a paramagnetic center which is close to that of a single electron. A well-known exception is a rare earth (RE) ion: it has valuable contribution to its magnetic moment from the orbital motion of its electrons due to the presence of strong spin-orbit coupling [21]. As a result, effective *g* factors of several RE ions under certain conditions exceed 10. Spin qubits based on RE ions [11-13] are advantageous as they would allow spin manipulations in low driving fields. Since the presence of magnetic anisotropy increases the number of experimentally controllable parameters that could in principle influence the decoherence times (namely, the directions of vectors $\boldsymbol{B}_0$ and $\boldsymbol{B}_1$), an appropriate choice of these parameters would enable one to increase the number of one-qubit operations.

In general, there are three non-equivalent directions related to the eigenvectors of anisotropic *g* tensor. In the present work, we consider the simplest case of axially symmetric background encountered when the local symmetry of the site occupied by the RE ion is tetragonal, trigonal or hexagonal. However, it is straightforward to modify the results for the case of lower symmetry (orthorhombic, monoclinic or triclinic crystal system). This paper is organized as follows: in Section I we develop a microscopic model of dipolar relaxation in transient regime and axially anisotropic crystal field. In Section II we illustrate our model by studying ROs in the concentration series of $CaWO_4:Nd^{3+}$ crystal.

**Section I. Driven dipolar relaxation in axially anisotropic crystal field**

Let us consider an ensemble of $N$ spins interacting with the external magnetic field $\boldsymbol{B} = \boldsymbol{B}_0 + 2\boldsymbol{B}_1 \cos\omega_0 t$ and with each other:

$$H = \mu_B \sum_{j=1}^{N} \hat{\mathbf{g}} \boldsymbol{B} \boldsymbol{S}^j + \sum_{j<k} \sum_{\alpha,\beta=x,y,z} d_{\alpha\beta}^{jk} S_{\alpha 1}^{j} S_{\beta 1}^{k} . \tag{2}$$

In the above Hamiltonian, $\boldsymbol{S}^j$ is the spin operator of the particle $j$, $\mu_B$ is the Bohr magneton, and $\hat{\mathbf{g}}$ is the axially symmetric *g* tensor written in its principal axes $x, y, z$

$$\hat{\mathbf{g}} = \begin{pmatrix} g_\perp & 0 & 0 \\ 0 & g_\perp & 0 \\ 0 & 0 & g_\parallel \end{pmatrix}. \tag{3}$$

Generally, the directions of $\boldsymbol{B}_0$ and $\boldsymbol{B}_1$ fields with respect to the axes $x, y, z$ are arbitrary. We choose the axes $x$ and $y$ so that $\boldsymbol{B}_0$ is in the $xz$ plane and at angle $\Theta$ from the $z$ axis. The direction of $\boldsymbol{B}_1$ is given by the direction cosines $h_x, h_y, h_z$:

$$\boldsymbol{B}_0 = \left(\boldsymbol{e}_x \sin\Theta + \boldsymbol{e}_z \cos\Theta\right) B_0, \quad \boldsymbol{B}_1 = \left(\boldsymbol{e}_x h_x + \boldsymbol{e}_y h_y + \boldsymbol{e}_z h_z\right) B_1. \tag{4}$$

We are going to apply two transformations in order to simplify the first term of the Hamiltonian (2). By the first transformation

$$S_{x0}^{j} = \left(g_{\parallel}S_{x}^{j}\cos\Theta - g_{\perp}S_{z}^{j}\sin\Theta\right)/g_{\Theta},$$
$$S_{y0}^{j} = S_{y}^{j}, \qquad (5)$$
$$S_{z0}^{j} = \left(g_{\parallel}S_{z}^{j}\cos\Theta + g_{\perp}S_{x}^{j}\sin\Theta\right)/g_{\Theta},$$

where $g_{\Theta} = \sqrt{\left(g_{\parallel}\cos\Theta\right)^{2} + \left(g_{\perp}\sin\Theta\right)^{2}}$, we diagonalize the interaction with the static field, $\hat{\mathbf{g}}\mathbf{B}_{0}\mathbf{S}^{j} = g_{\Theta}B_{0}S_{z0}^{j}$. Second transformation

$$S_{x1}^{j} = \left[\frac{g_{\parallel}g_{\perp}}{g_{\Theta}}\left(h_{x}\cos\Theta - h_{z}\sin\Theta\right)S_{x0}^{j} + g_{\perp}h_{y}S_{y0}^{j}\right]/g_{1},$$
$$S_{y1}^{j} = \left[\frac{g_{\parallel}g_{\perp}}{g_{\Theta}}\left(h_{x}\cos\Theta - h_{z}\sin\Theta\right)S_{y0}^{j} - g_{\perp}h_{y}S_{x0}^{j}\right]/g_{1}, \qquad (6)$$
$$S_{z1}^{j} = S_{z0}^{j}, \quad g_{1} = \frac{g_{\perp}}{g_{\Theta}}\sqrt{g_{\parallel}^{2}\left(h_{x}\cos\Theta - h_{z}\sin\Theta\right)^{2} + g_{\Theta}^{2}h_{y}^{2}}$$

is aimed at the interaction with the mw field, so that $\hat{\mathbf{g}}\mathbf{B}_{1}\mathbf{S}^{j} = g_{1}B_{1}S_{x1}^{j}$. We have neglected the term of the interaction $\sim B_{1}S_{z1}^{j}$ since it does not induce spin transitions. The Hamiltonian (2) can now be written as

$$H = \sum_{j=1}^{N}\left(\omega_{j}S_{z1}^{j} + 2\Omega_{R}S_{x1}^{j}\cos\omega_{0}t\right) + \sum_{j<k}\sum_{\alpha,\beta=x,y,z}D_{\alpha\beta}^{jk}S_{\alpha 1}^{j}S_{\beta 1}^{k}, \qquad (7)$$

where $\omega_{j} = \mu_{B}g_{\Theta}B_{0}$ is the Larmor frequency of the spin $j$ and $\Omega_{R} = \mu_{B}g_{1}B_{1}$ is the Rabi frequency ($\hbar = 1$). Defects of the crystal lattice bring random contributions to the crystal field resulting in the distribution of $g_{\perp}$ and $g_{\parallel}$. We assume here that the frequencies $\omega_{j}$ are distributed within the EPR line centered at $\omega_{0}$ and with the half-width $\sigma \ll \omega_{0}$. Usually $B_{1} \ll B_{0}$, and $\sigma$ can be as high as several $\Omega_{R}$. The Hamiltonian (7) now has the same form as in the isotropic case [17], except that $D_{\alpha\beta}^{jk}$ are certain linear combinations of the initial dipolar parameters $d_{\alpha\beta}^{jk}$. The Hamiltonian written in the rotating reference frame (RRF) defined by the unitary transformation $R = \exp\left(i\omega_{0}t\sum_{j=1}^{N}S_{z1}^{j}\right)$ is

$$H' = \sum_{j}\left(\varepsilon_{j}S_{z1}^{j} + \Omega_{R}S_{x1}^{j}\right) + \sum_{j<k}\left[D_{zz}^{jk}S_{z1}^{j}S_{z1}^{k} + \frac{D_{xx}^{jk} + D_{yy}^{jk}}{2}\left(S_{x1}^{j}S_{x1}^{k} + S_{y1}^{j}S_{y1}^{k}\right)\right], \qquad (8)$$

where $\varepsilon_{j} = \omega_{j} - \omega_{0}$ is the detuning of the spin $j$ from resonance frequency, and we have neglected time dependent terms of dipolar interaction not in resonance with any possible transition. Let us introduce local coordinate system $\tilde{x}^j, \tilde{y}^j, \tilde{z}^j$ in RRF associated with a given spin $j$ (see Fig. 1). The new spin operators that we will further mark with tilde are

$$\tilde{S}_x^j = \left(\varepsilon_j S_{z1}^j + \Omega_R S_{x1}^j\right)/\Omega_j, \quad \tilde{S}_y^j = S_{y1}^j, \quad \tilde{S}_z^j = \left(\Omega_R S_{z1}^j - \varepsilon_j S_{x1}^j\right)/\Omega_j, \quad \Omega_j = \sqrt{\varepsilon_j^2 + \Omega_R^2}, \qquad (9)$$

where $\Omega_j$ is the nutation frequency of the spin $j$ (note that $\Omega_j \geq \Omega_R$). The Hamiltonian (8) takes the following form:

$$H' = \sum_j \Omega_j \tilde{S}_x^j + \sum_{\substack{j<k \\ \alpha\beta}} \tilde{D}_{\alpha\beta}^{jk} \tilde{S}_\alpha^j \tilde{S}_\beta^k. \qquad (10)$$

It is clear that in the absence of dipolar couplings ($\tilde{D}_{\alpha\beta}^{jk} = 0$) the interaction of a given spin $j$ with the steady and mw magnetic fields would result in its precession with frequency $\Omega_j$ around the $\tilde{x}^j$ axis of RRF. The dipolar interactions introduce the correlations between the spin states, so that the dynamics of the spin $j$ would depend on positions and directions of the nearby spins $k$, which, in turn, are also influenced by their local spin background. The average strength of dipolar coupling in the dilute spin ensemble is determined by the dipolar half-width $\Delta\omega_d$ [22] which is linear in the spin concentration $C$. In the case of axially symmetric crystal field, one obtains [23]

$$\Delta\omega_d = \frac{4\pi^2 g_d^2 \mu_B^2 C}{9\sqrt{3}\hbar}, \quad g_d^2 = \frac{g_\perp^4 \sin^2\Theta + g_\parallel^4 \cos^2\Theta}{g_\perp^2 \sin^2\Theta + g_\parallel^2 \cos^2\Theta}. \qquad (11)$$

Usually, $C$ is small enough (<$10^{20}$ spins/cc), so that the condition $\Delta\omega_d \ll \sigma, \Omega_R$ is satisfied. We can neglect all terms in dipolar coupling except $\tilde{D}_{xx}^{jk} \tilde{S}_x^j \tilde{S}_x^k$, i.e. leave only a secular part with respect to the first term of Eq. (10). Indeed, the terms $\tilde{D}_{xy}^{jk} \tilde{S}_x^j \tilde{S}_y^k$ and $\tilde{D}_{xz}^{jk} \tilde{S}_x^j \tilde{S}_z^k$ that cause the transition of the spin $k$ with respect to $\tilde{x}^k$ axis would change the total energy by $\Omega_k$ and are unfavorable since $\tilde{D}_{\alpha\beta}^{jk} \ll \Omega_R$. The terms $\tilde{D}_{yy}^{jk} \tilde{S}_y^j \tilde{S}_y^k$ and $\tilde{D}_{yz}^{jk} \tilde{S}_y^j \tilde{S}_z^k$ related to mutual transitions of the spins $j$ and $k$ would change the energy by $\Omega_j \pm \Omega_k$ which are, on average, of the same order as either $\Omega_R$ or $\sigma$. If $t = 0$ is the moment the mw field was switched on, then, at any time $t > 0$, the magnetic moment of the spin ensemble is given by

$$\boldsymbol{M}(t) = \mu_B \hat{\mathbf{g}} \mathrm{Tr}\left\{ e^{-iH't} \rho e^{iH't} \sum_j \boldsymbol{S}^j \right\}. \qquad (12)$$

The initial density matrix $\rho$ can be written in the high-temperature approximation $\omega_0 \ll T$ generally valid even at liquid helium temperatures

$$\rho = \frac{1}{2^N} \prod_j \left(1 - \frac{\omega_0 S_{z1}^j}{T}\right). \qquad (13)$$

The calculation of the trace in Eq. (12) is best done in the basis $|m_j\rangle$, where $m_j = \pm 1$ are related to the eigenvalues $\pm 1/2$ of the spin operator $\tilde{S}_x^j$. Depending on the experimental pulse sequence, a certain projection of $\mathbf{M}(t)$ is detected. For example, the longitudinal (along the $z_1$ axis) component of the magnetization is

$$M_\parallel(t) = -\frac{g_\Theta \mu_B \omega_0 \Omega_R^2}{2^{N+2} T} \sum_j \Omega_j^{-2} \sum_{m_1, m_2, \ldots = \pm 1} \cos\left[\left(\Omega_j + \frac{1}{2}\sum_k{}' \tilde{D}_{xx}^{jk} m_k\right) t\right], \quad (14)$$

where the time-independent part of $M_\parallel$ is neglected, and the prime symbol in the last sum indicates that the term with $k = j$ is omitted. The argument of the cosine function has simple interpretation: the secular part of the interaction with the spin $k$ shifts the nutation frequency of the spin $j$ by $\tilde{D}_{xx}^{jk} m_k / 2$. Summation over all possible spin directions yields

$$M_\parallel(t) = \frac{M_0 \Omega_R^2}{N} \sum_j \Omega_j^{-2} \cos\Omega_j t \prod_k{}' \cos(\tilde{D}_{xx}^{jk} t / 2), \quad (15)$$

where $M_0 = -N g_\Theta \mu_B \omega_0 / 4T$ is the longitudinal magnetic moment at $t = 0$. The dipolar factor $\prod_k' \cos(\tilde{D}_{xx}^{jk} t/2)$ is responsible for the decay of ROs. As seen from Eq. (15), not all spins contribute equally to the ROs. Spins with large detuning ($\varepsilon_j \gg \Omega_R$) have negligible impact since $\Omega_R^2 / \Omega_j^2 \ll 1$. Spins with moderate detuning ($\varepsilon_j \sim \Omega_R$) represent valuable contribution during, roughly, the first period of oscillations, but after that they become dephased with respect to the resonant part of the ensemble. Since $\tilde{D}_{xx}^{jk} \ll \Omega_R$, the decay of ROs that is caused by the dipolar interactions reveals itself long after the first period, and we will further focus on the resonant spins ($\varepsilon_j \ll \Omega_R$). The subsequent calculations involve integration over random spin positions $\mathbf{r}_k$ within the crystal sample volume $V$ and over their frequency detunings $\varepsilon_k$ within the EPR line weighted with the spectral density $f(\varepsilon_k)$. We make the following assumptions: (a) the spin coordinates can be treated in the framework of the continuum approximation, i.e. regardless of the discrete periodic structure of the crystal lattice; (b) relative positions of any two spins, $\mathbf{r}_{jk} = \mathbf{r}_j - \mathbf{r}_k$, do not correlate with their detunings $\varepsilon_j$ and $\varepsilon_k$. These assumptions are the basics of the statistical method of line broadening [24] and are reasonable in the case of the spin concentrations less than 1 at. %. Thus, the averaging procedure starts as follows

$$\prod_k{}' \langle \cos(\tilde{D}_{xx}^{jk} t/2) \rangle_{\mathbf{r}_k, \omega_k} = \left\{\frac{1}{V} \int d\varepsilon_k f(\varepsilon_k) \int_V d^3 r_k \cos(\tilde{D}_{xx}^{jk} t / 2)\right\}^{N-1}. \quad (16)$$

In the macroscopic limit $N, V \to \infty$, while keeping $C = N/V = \text{const}$, one obtains

$$\prod_{k}{}' \left\langle \cos\left(\tilde{D}_{xx}^{jk}t/2\right)\right\rangle_{r_k,\omega_k} = \exp\left\{-C\int d\varepsilon_k f(\varepsilon_k)\int_\infty d^3 r_{jk}\left[1-\cos\left(\tilde{D}_{xx}^{jk}t/2\right)\right]\right\}. \qquad (17)$$

Integration over $r_{jk}$ gives (see Appendix A)

$$\int_\infty d^3 r_{jk}\left[1-\cos\left(\tilde{D}_{xx}^{jk}t/2\right)\right] = \frac{2\pi^2 \tilde{g}^2 \mu_B^2 \Omega_R t}{9\sqrt{3}\hbar \Omega_k}. \qquad (18)$$

This result has the same form as in the isotropic case [17], except that the isotropic *g*-factor is now substituted for the modified *g*-factor $\tilde{g}$ that depends on the ratio $g_\parallel/g_\perp$ and on the angle $\Theta$ (i.e. the direction of $\boldsymbol{B}_0$)

$$\tilde{g}^2 = \begin{cases} \dfrac{g_\parallel^2 g_\perp^2}{g_\Theta^2} G\!\left(\dfrac{g_\Theta^2}{g_\parallel^2}\right), & g_\parallel > g_\perp, \\[1em] g_\perp^2 G\!\left(\dfrac{g_\parallel^2}{g_\Theta^2}\right), & g_\parallel < g_\perp. \end{cases} \qquad (19)$$

Function $G(\xi)$ is shown in Fig. 2. Combining this result with Eq. (15), we obtain the longitudinal magnetization

$$M_\parallel(t) = M_0 \Omega_R^2 \exp(-\Gamma_d t)\int d\varepsilon f(\varepsilon)\frac{\cos\sqrt{\Omega_R^2+\varepsilon^2}\,t}{\Omega_R^2+\varepsilon^2}, \qquad (20)$$

or, in much the same way, the transverse (along the $y_1$ axis) component of the magnetization

$$M_\perp(t) = M_1 \Omega_R \exp(-\Gamma_d t)\int d\varepsilon f(\varepsilon)\frac{\sin\sqrt{\Omega_R^2+\varepsilon^2}\,t}{\sqrt{\Omega_R^2+\varepsilon^2}}, \qquad (21)$$

where $M_1 = -Ng_\Theta \mu_B \omega_0/4T$, and $\Gamma_d$ is the dipolar-induced decay rate

$$\Gamma_d = \frac{1}{2}\Delta\tilde{\omega}_d \Omega_R \int \frac{f(\varepsilon)d\varepsilon}{\sqrt{\Omega_R^2+\varepsilon^2}}. \qquad (22)$$

The modified dipolar half-width $\Delta\tilde{\omega}_d$ has the same form as in Eq. (11), but with $\tilde{g}$ instead of $g_d$. Function $G(\xi)$ can be replaced by unity in approximate calculation since $0.82 < G(\xi) \le 1$. A certain choice of $\Theta$ would minimize $\tilde{g}^2$ that enters the decay rate $\Gamma_d$ and, consequently, increase the number of coherent oscillations $n = \Omega_R/2\pi\Gamma_d$. This increase is considerable only when $g_\parallel$ is larger than $g_\perp$. In this case, the favorable direction of the static magnetic field would be close to *z* axis ($\Theta = 0$), with $\tilde{g}_{\min} \approx g_\perp$. In the case when $g_\parallel < g_\perp$, only a small deviation of $\tilde{g}$ from the in-plane *g*-factor $g_\perp$ is expected. Thus, $\tilde{g}_{\min} \approx g_\perp$ regardless of the ratio $g_\parallel/g_\perp$. Note that *n* indirectly depends on the direction and strength of the mw field $\boldsymbol{B}_1$ since the latter determines the Rabi frequency $\Omega_R$. The integration over $\varepsilon$ in Eqs. (20), (22) is straightforward if one knows the exact EPR lineshape function $f(\varepsilon)$. There are, however, two important limiting cases when the final result can be expressed in general form:

(a) Narrow line $\sigma \ll \Omega_R$. The lineshape function can be approximated by Dirac delta function $f(\varepsilon) = \delta(\varepsilon)$, all spins have their nutation frequencies equal to $\Omega_R$, and we obtain

$$\begin{aligned} \Gamma_d &= \Delta\tilde{\omega}_d/2, \\ M_\parallel(t) &= M_0 e^{-\Delta\tilde{\omega}_d t/2} \cos\Omega_R t, \\ M_\perp(t) &= M_1 e^{-\Delta\tilde{\omega}_d t/2} \sin\Omega_R t. \end{aligned} \qquad (23)$$

The decay rate reaches its highest value (a half of the modified dipolar half-width) and does not depend on the Rabi frequency.

(b) Broad line $\sigma \gg \Omega_R$. Since now only the central part of the EPR line is exited, the lineshape function can be replaced by its resonance value $f(0)$, and

$$\begin{aligned} \Gamma_d &= \Delta\tilde{\omega}_d f(0)\Omega_R \ln\frac{\sigma+\sqrt{\sigma^2+\Omega_R^2}}{\Omega_R}, \\ M_\parallel(t) &= M_0 \pi f(0)\Omega_R e^{-\Gamma_d t} j_0(\Omega_R t), \\ M_\perp(t) &= M_1 \pi f(0)\Omega_R e^{-\Gamma_d t} J_0(\Omega_R t) \quad (\sigma t \gg 1). \end{aligned} \qquad (24)$$

Here, $J_0(\xi)$ is the Bessel function of the first kind, and $j_0(\xi) = \int_\xi^\infty J_0(\zeta)d\zeta$. In most cases, these functions can be approximated by the slowly decaying cosine

$$j_0(\xi > 1) \approx J_0(\xi + \pi/2) \approx \sqrt{2}\left(1+(\pi\xi)^2\right)^{-1/4} \cos(\xi+\pi/4). \qquad (25)$$

While the above asymptotic relations are valid for arbitrary symmetric $f(\varepsilon)$, exact results can be derived irrespective of $\Omega_R/\sigma$ ratio in the two frequently encountered cases of Gaussian $f^{(G)}(\varepsilon) = (2\pi\sigma^2)^{-1/2} \exp\left[-\varepsilon^2/2\sigma^2\right]$ and Lorentzian $f^{(L)}(\varepsilon) = \sigma/\left[\pi(\varepsilon^2+\sigma^2)\right]$ lineshapes:

$$\begin{aligned} \Gamma_d^{(G)} &= \frac{\Omega_R \Delta\tilde{\omega}_d}{2\sigma\sqrt{2\pi}} \exp\left[\left(\frac{\Omega_R}{2\sigma}\right)^2\right] K_0\left[\left(\frac{\Omega_R}{2\sigma}\right)^2\right], \\ \Gamma_d^{(L)} &= \begin{cases} \dfrac{\Omega_R \Delta\tilde{\omega}_d}{\pi\sqrt{\sigma^2-\Omega_R^2}} \ln\dfrac{\sigma+\sqrt{\sigma^2-\Omega_R^2}}{\Omega_R}, & \dfrac{\Omega_R}{\sigma} < 1, \\ \dfrac{\Omega_R \Delta\tilde{\omega}_d}{\pi\sqrt{\Omega_R^2-\sigma^2}} \arccos\left(\dfrac{\sigma}{\Omega_R}\right), & \dfrac{\Omega_R}{\sigma} > 1, \end{cases} \end{aligned} \qquad (26)$$

where $K_0(\xi)$ is the modified Bessel function of the second kind. As shown in Fig. 3, $\Gamma_d$ grows monotonously with the ratio $\Omega_R/\sigma$ and tends to its limiting value $\Delta\tilde{\omega}_d/2$ at high Rabi frequencies. Let us now draw comparison between our results and the predictions of the phenomenological models [18, 19]. If the range of $\Omega_R/\sigma$ is small enough, the relaxation rate can indeed be approximated by the linear dependence (1) (see the dashed line in Fig. 3). However, this dependence is not universal since the coefficients $\alpha, \beta$ depend on the point of the curve through which a tangent line is drawn. It is clear that on a wider range of Rabi frequencies the approximation (1) becomes incorrect. Our experimental results presented in Sec. II confirm the nonlinearity of $\Gamma_d(\Omega_R)$.

**Section II. Rabi oscillations in CaWO$_4$:Nd$^{3+}$ crystal**

CaWO$_4$ single crystal has scheelite structure with lattice constants $a$ = 5.243 Å, $c$ = 11.374 Å [25]. Nd$^{3+}$ ions substitute for Ca$^{2+}$ ions in the host crystal at sites with $S_4$ point symmetry. The samples of Nd-doped CaWO$_4$ single crystal were grown by Czochralski method in Magnetic Resonance Laboratory of Kazan Federal University by N. A. Karpov. Experimental data were acquired by means of Bruker Elexsys 580/680 X-band spectrometer working at mw frequency $\omega_0/2\pi$ =9.7 GHz and at temperature $T$ = 6 K. Actual concentration of neodymium ions in each sample ($C$ = 4.00·10$^{17}$÷1.04·10$^{20}$ ions per cc, see Table I) was determined by comparative measurement of the EPR line intensities with respect to the reference sample of CaF$_2$:Er$^{3+}$ (0.28 at. %). Continuous-wave EPR spectrum shown in Fig. 4 contained an intense central peak arising from even Nd isotopes with nuclear spin $I$ = 0 (natural abundance 79.5%), and a number of hyperfine satellites coming from $^{143}$Nd ($I$ = 7/2, 12.2%) and $^{145}$Nd ($I$ = 7/2, 8.3%). The lines had nearly Lorentzian lineshapes and almost equal half-widths $\sigma$ that varied with $C$ and the sample orientation. Our crystal field calculations, as well as the experimental data, are in agreement with the literature $g$-factor values $g_\parallel$ =2.034 and $g_\perp$ =2.528 [26].

The measurements described below were taken at the central peak and at certain $^{143}$Nd and $^{145}$Nd satellites (see the arrows in Fig. 4). The orientation of the crystal sample in the mw resonator was chosen to be $\boldsymbol{B}_0 \perp c$, $\boldsymbol{B}_1 \parallel c$, with the exception of the sample no. 4, where both $\boldsymbol{B}_0$ and $\boldsymbol{B}_1$ were perpendicular to the crystal $c$ axis. First of all, spin-lattice relaxation times $T_1$ and phase memory times $T_2$ were obtained for each sample in the concentration series (see Table I). The length of $\pi/2$ pulse was 8 ns in all $T_1$ and $T_2$ measurements. Because of the role of random electric fields, $\sigma$ depended on the exact orientation of $\boldsymbol{B}_0$ in $ab$ plane, with minima and maxima of $\sigma$ at certain angles [27-28]. For comparison reasons, all the data presented below were recorded at the minima of $\sigma$. At the maxima, $T_2$ and $\tau_R$ were several percent longer, while $T_1$ showed no visible variation. As for the Rabi times, this result is much expected since in the latter case the mw pulse affects less number of Nd ions. Similar increase of $T_2$ times and their dependence on the isotopic concentration are in accordance with the theoretical estimations that indicate instantaneous diffusion and spectral diffusion as dominant contributions into the phase relaxation in $CaWO_4$:$Nd^{3+}$ crystal [29]. The spin-lattice relaxation times for the first three samples were in the range $T_1 = 15 \div 25$ ms and did not vary with the isotopic concentration; these results are consistent with the literature data [30, 31], where direct and Raman processes are singled out as being the dominant contributions. However, we cannot give a direct account of the abrupt decrease of $T_1$ in the last sample with the highest neodymium concentration. This change may arise as the result of local deformation of the crystal lattice near the paramagnetic impurity and subsequent perturbation of the vibrational spectrum of the crystal, which is more pronounced at higher $C$.

Each data point of the ROs was obtained after the pulse sequence shown in Fig. 5, where the transient pulse was followed by the spin-echo detection sequence which finally gave the longitudinal component of the magnetization $M_\parallel$. Some of the recorded ROs are presented in Fig. 6 and Fig. 7. $M_\parallel(t)$ were calculated in the most general way according to the Section I as

$$M_\parallel(t) = A(t)\cos(\Omega_R t + \varphi),$$
$$A(t) \sim \left[1+(\beta_{B1}\Omega_R t)^2\right]^{-3/4} \left[1+(\pi\Omega_R t)^2\right]^{-\mu} \exp(-\kappa\Gamma_d t). \quad (27)$$

A decay factor $\left[1+(\beta_{B1}\Omega_R t)^2\right]^{-3/4}$ was added to the amplitude $A(t)$ in order to account for the spatial distribution of $B_1$ in the mw resonator [9]. The corresponding decay rate is linear in $\Omega_R$: $\Gamma_{B1} = \beta_{B1}\Omega_R$. However, in contrast to the dipolar contribution $\exp(-\kappa\Gamma_d t)$, the $B_1$-type decay is determined by the slowly reducing rational function. The inhomogeneity parameter $\beta_{B1}$ represents relative decrease of $\Omega_R(r)$ at the sample edges with respect to its maximal value $\Omega_R = \Omega_R(0)$ at the center of the cavity. In most cases, $\beta_{B1} \leq 0.1$, so this effect can be neglected in $T_1$ and $T_2$ measurements, where short pulses with the lengths less than the Rabi period are used. For a small sample with the dimensions $l_x \times l_y \times l_z$ placed at the center of $TE_{011}$ cylindrical resonant cavity of radius $R$ and length $L$ [9]

$$\beta_{B1} = v_{01}^2\left(l_x^2+l_y^2\right)/16R^2 + \pi^2 l_z^2/8L^2, \quad v_{01} = 3.832. \quad (28)$$

The parameters $\beta_{B1}$ that were found best to describe the experimental data in the samples no. 1-4 are presented in Table I. They are very close to the value $\beta_{B1} = 0.05$ for the 3 mm sample that was estimated according to Eq. (28). The two other parameters, $0 \leq \varphi \leq \pi/4$ and $0 \leq \mu \leq 1/4$, are determined by the ratio $\Omega_R/\sigma$. As follows from Eqs. (23) and (24),

$$\begin{aligned} \varphi = 0, \quad \mu = 0 \quad (\Omega_R/\sigma < 1), \\ \varphi = \pi/4, \quad \mu = 1/4 \quad (\Omega_R/\sigma > 1). \end{aligned} \quad (29)$$

The parameter $\kappa$ that was introduced into the exponent in Eq. (27) is the ratio of neodymium ions corresponding to the given EPR line to the total number of $Nd^{3+}$ ions in the crystal sample: $\kappa = 0.795$, $0.015$ and $0.01$ for the central line, $^{143}Nd$ and $^{145}Nd$ satellites, respectively.

Rabi rates $\tau_R^{-1}$ as functions of $\Omega_R$ collected from all four samples are represented by symbols in Fig. 8. They are in excellent agreement with the calculated dependences (solid and dashed lines). Rabi rates $\tau_R^{-1}$ always grew monotonously with $\Omega_R$, i.e. with the strength of mw field. For the samples no. 1 and no. 2 with lower $C$, the dependence $\tau_R^{-1}(\Omega_R)$ was almost linear, meaning that the dominant contribution came from $B_1$ inhomogeneity. This also accounts for the fact that there is only a small difference between the nutation signal of different neodymium isotopes in these samples. On the contrary, $\tau_R^{-1}(\Omega_R)$ of the sample no. 4 was nonlinear, indicating the domination of dipolar contribution; the decay rates of $^{143}Nd$ and $^{145}Nd$ isotopes were much smaller than the ones of the central line (see Fig. 7 and the dashed line in Fig. 8). Note that in our calculations we did not account for the dynamics of the nuclear spin of Nd ion. The hyperfine interaction would result in the renormalization of $\Omega_R$ and of the dipolar interaction parameter $\tilde{D}_{xx}^{jk}$, thus changing the decay rate, especially in the case when the Larmor frequency of the hyperfine satellite differs substantially from that of the central line. The corresponding corrections are of order $|Am_I|/\omega_0$, where $A$ is the hyperfine coupling parameter, $m_I$ is the nuclear spin projection (see Appendix B). Our experimental data were obtained at the closest $^{143}Nd$ and $^{145}Nd$ satellites corresponding to $m_I = -1/2$; in this case, $|Am_I|/\omega_0 \sim 0.01$, and the hyperfine correction to $\tau_R$ is negligible.

As was expected, the longest coherence times were obtained for the sample no. 1. There we observed $\tau_R$ up to 1 µs and over 50 visible periods of ROs. That long-lasting transient coherence permitted us to detect an interesting phenomenon. In Fig. 9 one can see the amplitude modulation resulting from the interference of the signals that come from different parts of the crystal sample. The arrow shows the dip at $t_0 = 0.55$ µs which is the first point of destructive interference. Roughly, one expects this dip to occur when the phases of the oscillations at the center ($\varphi_c$) and at the edge ($\varphi_e$) of the sample differ by $\pi$. It follows from the calculations presented in our previous paper [9] that $\varphi_c = \Omega_R t$ and $\varphi_e = (1 + \beta_{B1})\Omega_R t$. This gives us an estimated value $t_0 = \pi/\beta_{B1}\Omega_R = 0.67$ μs which is in reasonable agreement with the experimental one.

**Conclusions**

To the best of our knowledge, this work represents the first quantitative description of ROs of paramagnetic impurity ions in the anisotropic crystal field. We developed a microscopic theory of dipolar relaxation in transient regime that contained no phenomenological parameters and, in contrast to existing phenomenological models, predicted nonlinear dependence of the decay rate on the Rabi frequency. In addition, we accomplished the first experimental study of ROs in the concentration series of Nd:CaWO$_4$ single crystals. The obtained experimental data for the whole range of spin concentrations, the strengths of the mw field and isotopic numbers of Nd ions are in excellent agreement with our ab initio calculations.

At last, let us discuss the relation between the spin coherence times $T_2$ and $\tau_R$. In quantum computation processing, it is advantageous to increase both these quantities in order to obtain higher number of qubit operations. Generally, the ratio $T_2/\tau_R$ depends on the spin concentration, on the field inhomogeneity inside the crystal sample, and on the strength of the mw field during the transient pulse and the spin-echo sequence. Under our experimental conditions, we found $T_2/\tau_R = 1 \div 300$. The longest $\tau_R \sim 1$ μs were observed in the sample with the lowest spin concentration. It was possible to increase the Rabi times by using smaller crystal samples in order to reduce the inhomogeneity of $B_1$, but with the substantial loss of the signal intensity.

E.I.B. acknowledges the support of Dynasty foundation and the Russian Government Program of Competitive Growth of Kazan Federal University.

**Appendix A. Calculation of the dipolar factor**

Here we calculate the integral in the left part of the Eq. (18)

$$J = \int_\infty d^3 r_{jk} \left[1 - \cos\left(\tilde{D}_{xx}^{jk} t/2\right)\right]. \tag{30}$$

First of all, we need to express the effective dipolar coupling $\tilde{D}_{xx}^{jk}$ through the initial dipolar parameters $d_{\alpha\beta}^{jk}$. This can be done using Eqs. (2)-(10), and $d_{\alpha\beta}^{jk}$ are defined by the following relation

$$\sum_{\alpha,\beta=x,y,z} d_{\alpha\beta}^{jk} S_\alpha^j S_\beta^k = \frac{\mu_B^2}{\hbar r_{jk}^3} \left\{ \left(\hat{g}S^j \cdot \hat{g}S^k\right) - \frac{3\left(\hat{g}S^j \cdot r_{jk}\right)\left(\hat{g}S^k \cdot r_{jk}\right)}{r_{jk}^2} \right\}. \tag{31}$$

Thus we obtain

$$\tilde{D}_{xx}^{jk} = -\frac{\left(g_\parallel^2 + g_\Theta^2\right)g_\perp^2 \mu_B^2 \Omega_R}{4\hbar g_\Theta^2 r^3 \Omega_k}\left\{1 - 3\cos^2\theta + 3\frac{g_\parallel^2 - g_\Theta^2}{g_\parallel^2 + g_\Theta^2}\sin^2\theta\cos 2\varphi\right\}, \quad (32)$$

where $r, \theta, \varphi$ are spherical coordinates of the vector $\mathbf{r}_{jk}$. Integration over $r$ yields

$$J = \frac{\pi\left(g_\parallel^2 + g_\Theta^2\right)g_\perp^2 \mu_B^2 \Omega_R t}{6\hbar g_\Theta^2 \Omega_k}\int_0^{\pi/2} d\varphi \int_0^1 d\xi \left|1 - 3\xi^2 + 3\delta\left(1 - \xi^2\right)\cos 2\varphi\right| \quad (33)$$

where $\delta = \left|g_\parallel^2 - g_\Theta^2\right|/\left(g_\parallel^2 + g_\Theta^2\right)$, and finally

$$J = \frac{2\pi^2 \tilde{g}^2 \mu_B^2 \Omega_R t}{9\sqrt{3}\hbar \Omega_k}, \quad \tilde{g}^2 = \begin{cases} \dfrac{g_\parallel^2 g_\perp^2}{g_\Theta^2}G\left(\dfrac{g_\Theta^2}{g_\parallel^2}\right), & g_\parallel > g_\perp, \\ g_\perp^2 G\left(\dfrac{g_\parallel^2}{g_\Theta^2}\right), & g_\parallel < g_\perp. \end{cases} \quad (34)$$

Function $G(\xi)$ (Fig. 2) is expressed through the complete elliptic integrals $K(\zeta)$ and $E(\zeta)$ as

$$G(\xi) = \begin{cases} \dfrac{\sqrt{1-2\xi}}{\pi}\left\{3E\left[\dfrac{\xi(2-\xi)}{2\alpha-1}\right] - (1+\alpha)K\left[\dfrac{\xi(2-\xi)}{2\alpha-1}\right]\right\}, & 0 \le \xi < \dfrac{1}{2}, \\ G(1-\xi), & \dfrac{1}{2} < \xi \le 1. \end{cases} \quad (35)$$

**Appendix B. An account of the hyperfine interaction**

The hyperfine interaction (hfi) that is present in the case of $^{143}$Nd and $^{145}$Nd ions was not included into the Hamiltonian (2). Let us now estimate if it has any influence on $\tau_R$ under our experimental conditions. The hfi of a given neodymium ion $j$ with its nuclear spin $I$ (index $j$ is omitted below for simplicity) is

$$H_{hfi} = A_\parallel S_z I_z + A_\perp \left(S_x I_x + S_y I_y\right), \quad (36)$$

where $A_\parallel = g_\parallel A/g_J$ and $A_\perp = g_\perp A/g_J$, $g_J$ is the Lande g-factor, $A/2\pi = -220$ MHz and $A/2\pi = -137$ MHz represent the hyperfine coupling constants for the isotopes $^{143}$Nd and $^{145}$Nd, respectively [21]. In the electronic coordinate system (6) the above interaction takes the form

$$H_{hfi} = \sum_{\alpha,\beta=x,y,z} A_{\alpha\beta} S_{\alpha 1} I_{\beta 2}, \quad (37)$$

where $A_{\alpha\beta}$ are certain linear combinations of $A_\parallel$ and $A_\perp$. Index "2" in the operator $I_{\beta 2}$ denotes a specific rotation of the nuclear coordinate system that is applied in order to exclude the terms with $A_{zx}$ and $A_{zy}$. Since (i) the relaxation time of the nuclear states in magnetically diluted crystals is usually much longer than $\tau_R$, (ii) $|A| \ll \omega_0$, and (iii) the interaction energy of the nuclear spin with the external magnetic field is negligible with respect to $A$, the projection $m_I$ of the nuclear spin along the $z_2$ axis represents a good quantum number. Indeed, the terns of the hfi $A_{xx}S_{x1}I_{x2}$, $A_{xy}S_{x1}I_{y2}$, etc. that are responsible for the nuclear transitions are ineffective since they also change the electron spin energy by $\omega_0$. We can now replace $I_{z2}$ with $m_I$ and neglect the terms with $I_{x2}$ and $I_{y2}$ that mix different nuclear spin states:

$$H_{hfi} = A_{zz}S_{z1}m_I + \left(A_{xz}S_{x1} + A_{yz}S_{y1}\right)m_I. \tag{38}$$

This interaction should be added to each $j$ term of the Hamiltonian (7). The first part in the right-hand side of Eq. (38) gives a shift $A_{zz}m_I$ of the spin Larmor frequency $\omega$ that results in the complex hyperfine structure which is clearly visible in the EPR spectrum (see Fig. 4). Since the second part of Eq. (38) contains no time-dependent terms $\sim e^{\pm i\omega_0 t}$, it does not shift $\Omega_R$ directly. Instead, it slightly tilts the quantization axis of the electron spin from $z_1$ direction and finally yields rather small (~ a factor of $|Am_I|/\omega_0$) corrections to $\omega$, $\Omega_R$ and to the dipolar coupling parameters. The full expressions with explicit dependences on $\mathbf{B}_0$ and $\mathbf{B}_1$ direction cosines are rather cumbersome and need not be given here. Note that our experimental data were obtained at the central line ($I = 0$) and at the closest $^{143}$Nd and $^{145}$Nd satellites ($I = 7/2$, $m_I = -1/2$); in the latter case, $|Am_I|/\omega_0 \sim 0.01$, and the hyperfine correction to $\tau_R$ is negligible. Even for the most distant satellites with $m_I = \pm 7/2$ this correction is rather small. However, at radio frequencies ($\omega_0/2\pi \sim 300$ MHz) hfi would definitely play an important role. The theory in this specific case cannot be based on the perturbation approach and lies beyond the scope of the present work.

**Tables**

TABLE I. Concentration $C$ of $Nd^{3+}$ ions, half-width $\sigma$, inhomogeneity parameter $\beta_{B1}$ and relaxation times $T_1$ and $T_2$ in the crystal samples no. 1-4.

| Sample no. | | 1 | 2 | 3 | 4 |
|---|---|---|---|---|---|
| $C$ | ions per cc | $4.00 \cdot 10^{17}$ | $1.29 \cdot 10^{18}$ | $6.64 \cdot 10^{18}$ | $1.04 \cdot 10^{20}$ |
| | atomic % | 0.0031 | 0.010 | 0.052 | 0.81 |
| $\sigma/2\pi$, MHz | | 3.5 | 3.4 | 5.0 | 47 |
| $\beta_{B1}$ | | 0.05 | 0.05 | 0.05 | 0.06 |
| $T_1$, ms | Central line | 23 | 23 | 15 | 0.1 |
| | $^{143}$Nd | 23 | 24 | 16 | - |
| | $^{145}$Nd | 23 | 25 | 15 | - |
| $T_2$, μs | Central line | 2.5 | 1.0 | 0.4 | 0.14 |
| | $^{143}$Nd | 80 | 25 | 3.5 | 0.25 |
| | $^{145}$Nd | 100 | 29 | 4.2 | 0.16 |

**Figure captions**

FIG. 1. (Color online) Local coordinates in the rotating reference frame associated with a given spin $j$ (see text).

FIG. 2. (Color online) Function $G(\xi)$ that enters the modified g-factor $\tilde{g}$ (19).

FIG. 3. (Color online) The dipolar-induced decay rate expressed in units of $\Delta\tilde{\omega}_d$ as function of the ratio $\Omega_R/\sigma$. Two thick lines represent the cases of Gaussian and Lorentzian lineshapes (26), while two thin lines are their asymptotic approximations calculated according to Eq. (24). Dashed line represents linear approximation (1) in the range $0.05\sigma \leq \Omega_R \leq 0.3\sigma$.

FIG. 4. (Color online) EPR spectrum of the sample no. 1. $\boldsymbol{B}_0 \perp c$, $T = 15$ K.

FIG. 5. (Color online) The pulse sequence that was used for acquisition of ROs.

FIG. 6. (Color online) ROs in the sample no. 3 recorded at different strengths of mw field (circles). Longitudinal magnetization $M_\parallel(t)$ (solid line) and its envelope (dashed line) were calculated according to Eq. (27). (a) $\Omega_R/2\pi = 1.8$ MHz; (b) $\Omega_R/2\pi = 4.5$ MHz; (c) $\Omega_R/2\pi = 8.2$ MHz.

FIG. 7. (Color online) ROs in the sample no. 4 (circles). Longitudinal magnetization $M_\parallel(t)$ (solid line) and its envelope (dashed line) were calculated according to Eq. (27). $\Omega_R/2\pi = 6$ MHz. (a) central line; (b) $^{143}$Nd.

FIG. 8. (Color online) Measured (symbols) and calculated (curves) decay rates of ROs $\tau_R^{-1}$ as functions of Rabi frequencies $\Omega_R/2\pi$ in the samples no. 1-4. Squares, triangles and circles correspond to the data recorded at the central line, $^{143}$Nd and $^{145}$Nd satellites, respectively.

FIG. 9. (Color online) Amplitude modulation of ROs resulting from the interference of the signals coming from different parts of the sample no. 1. The arrow points on the dip located near the time point $t = \pi/\beta_{B1}\Omega_R$.

**Figures**

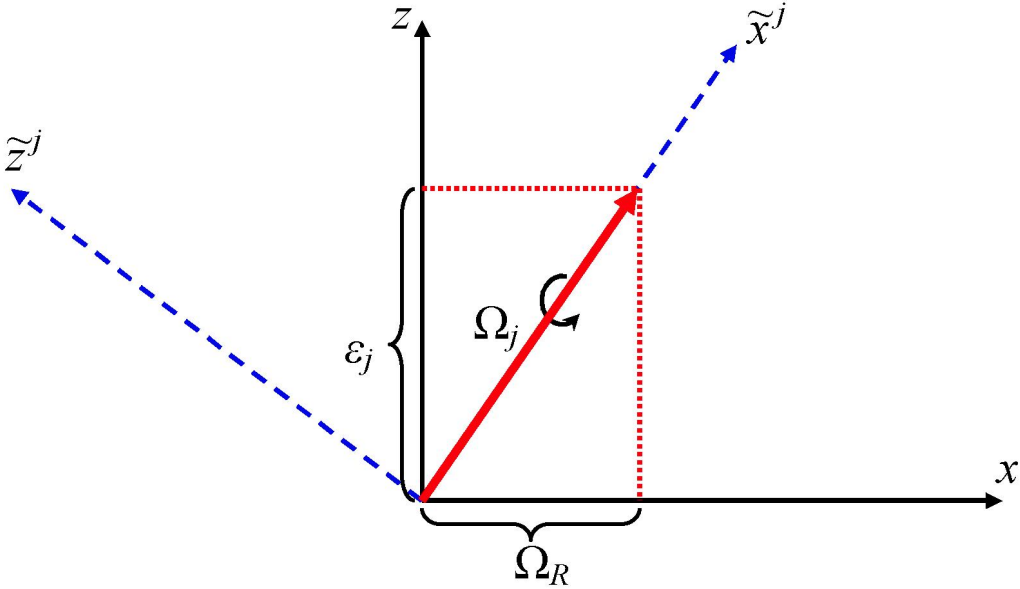

Figure 1

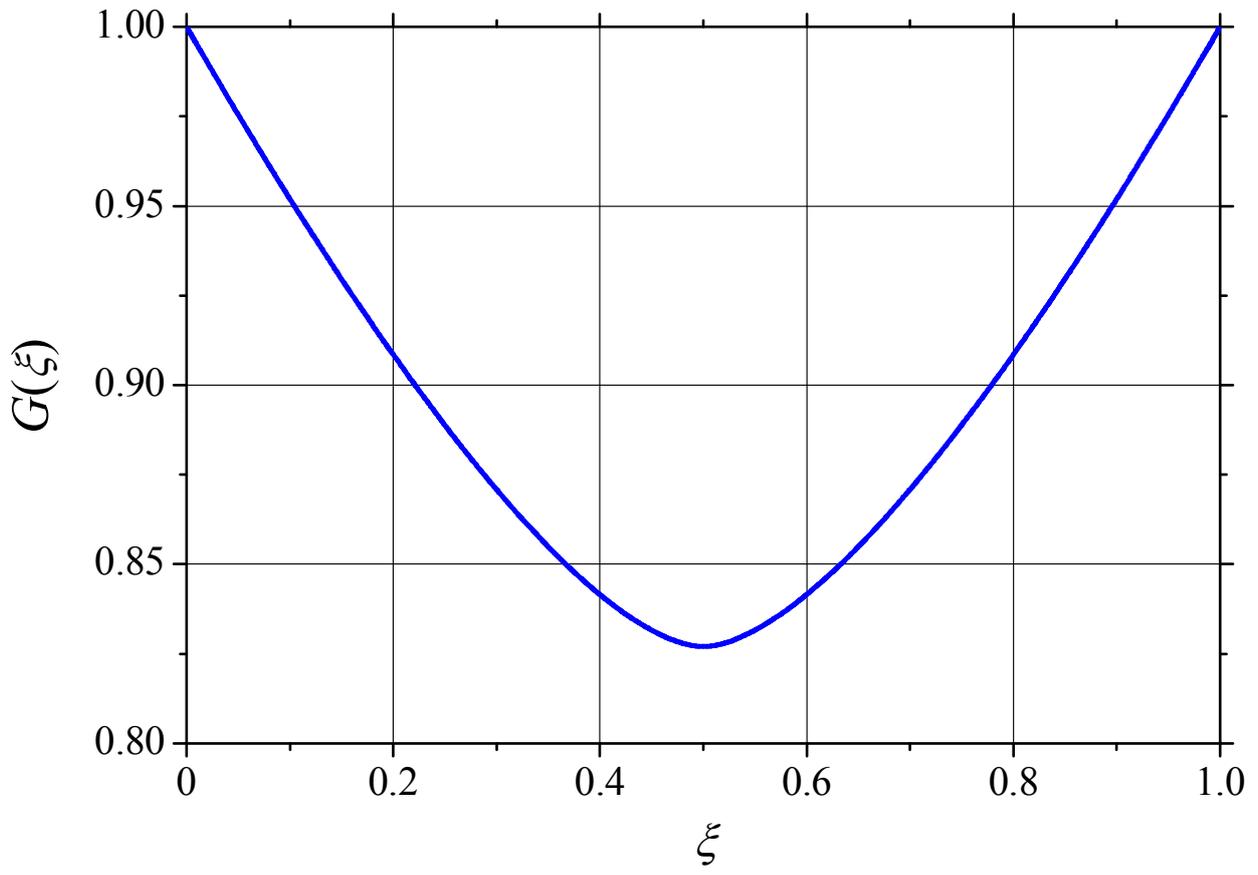

Figure 2

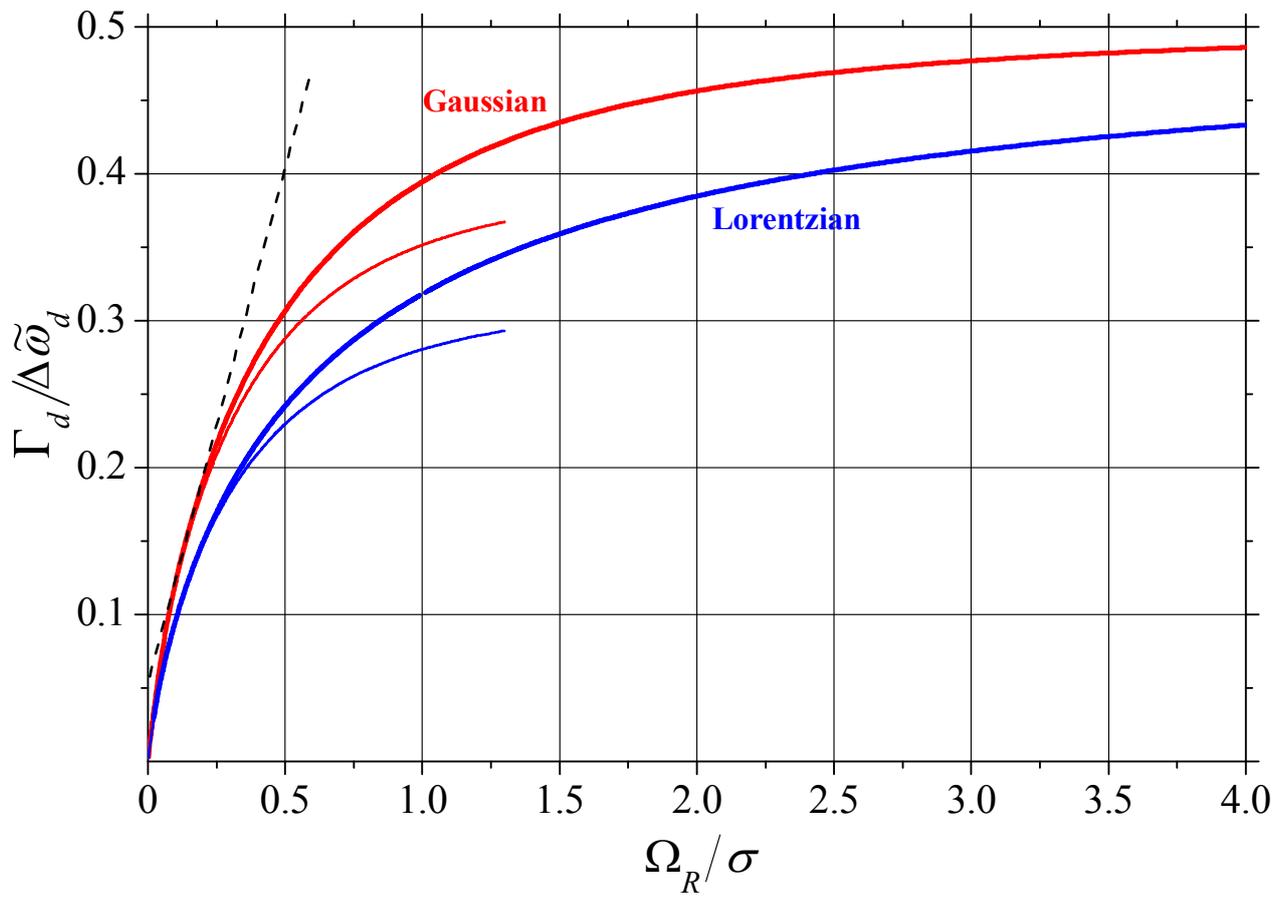

Figure 3

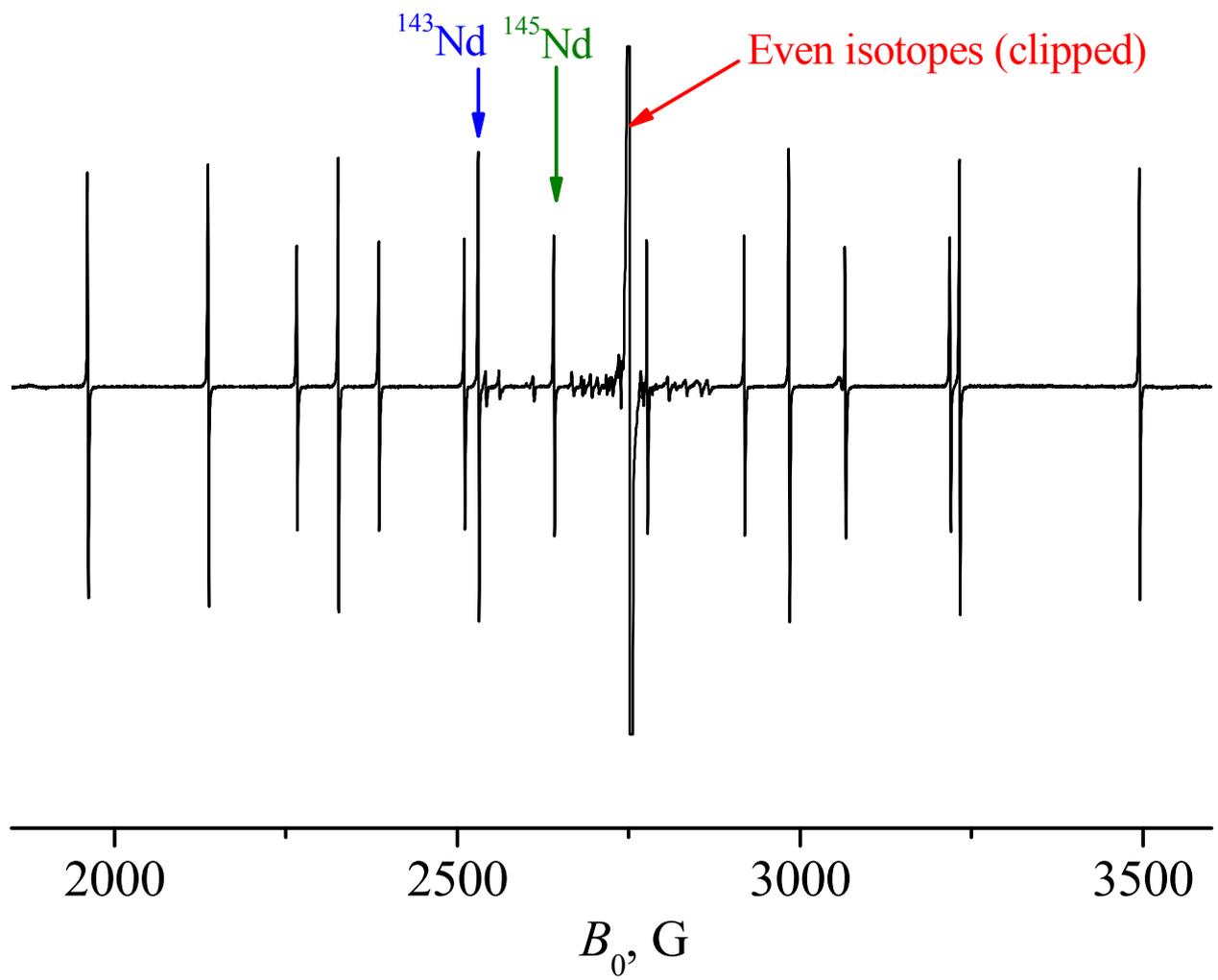

Figure 4

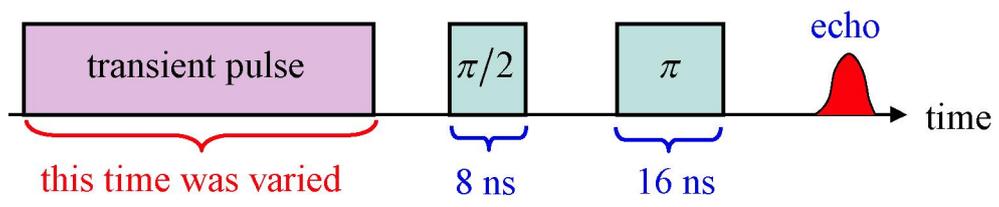

Figure 5

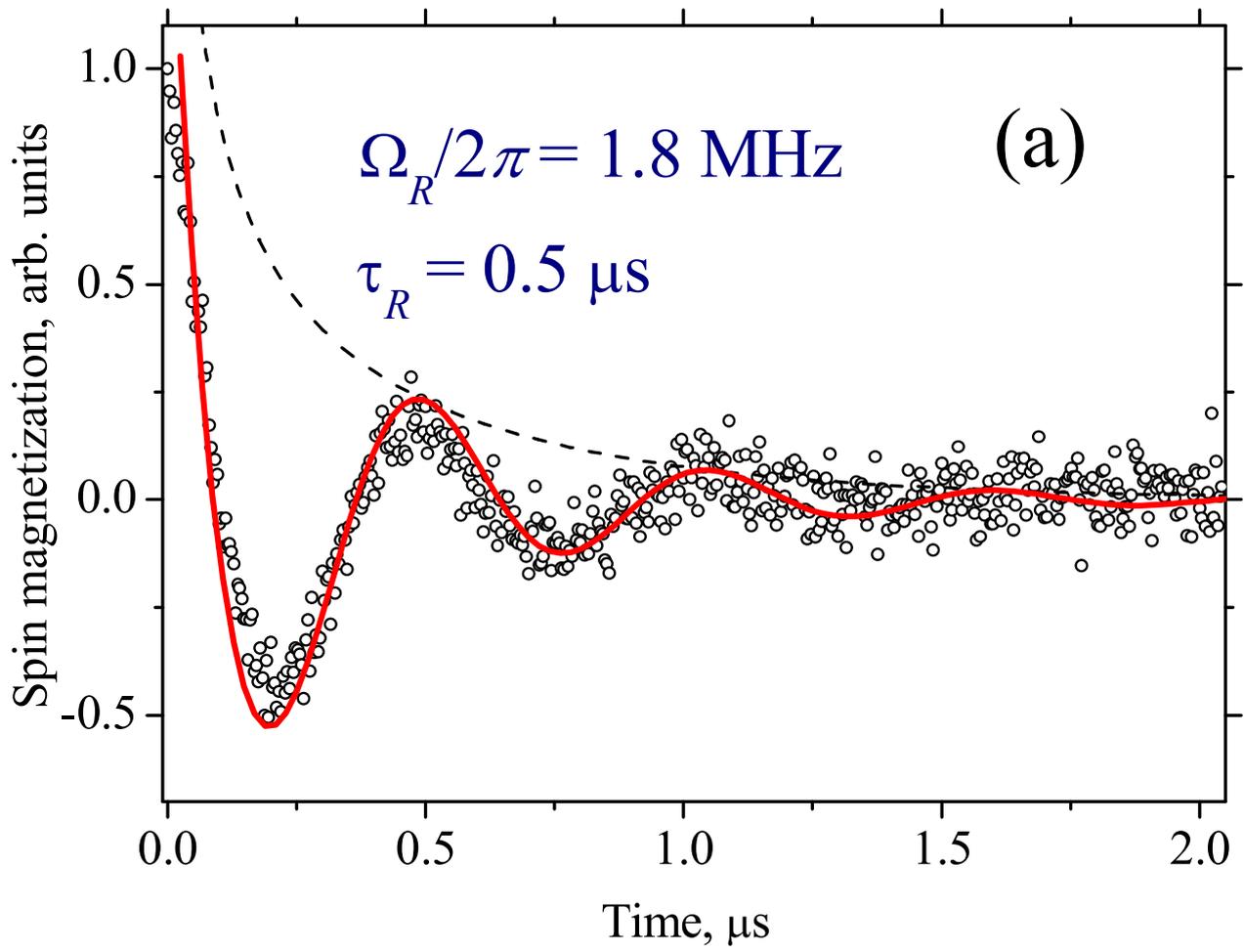

Figure 6a

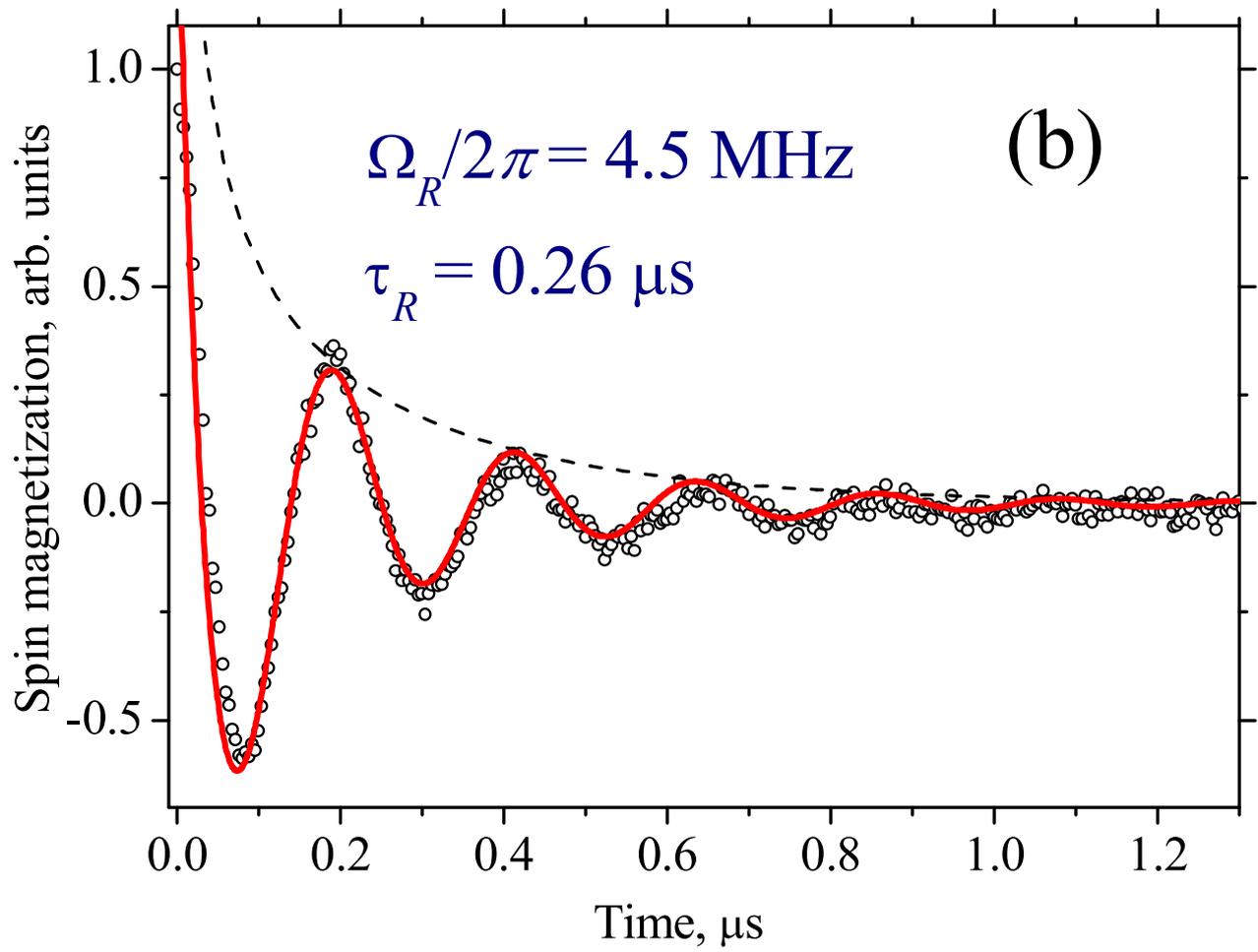

Figure 6b

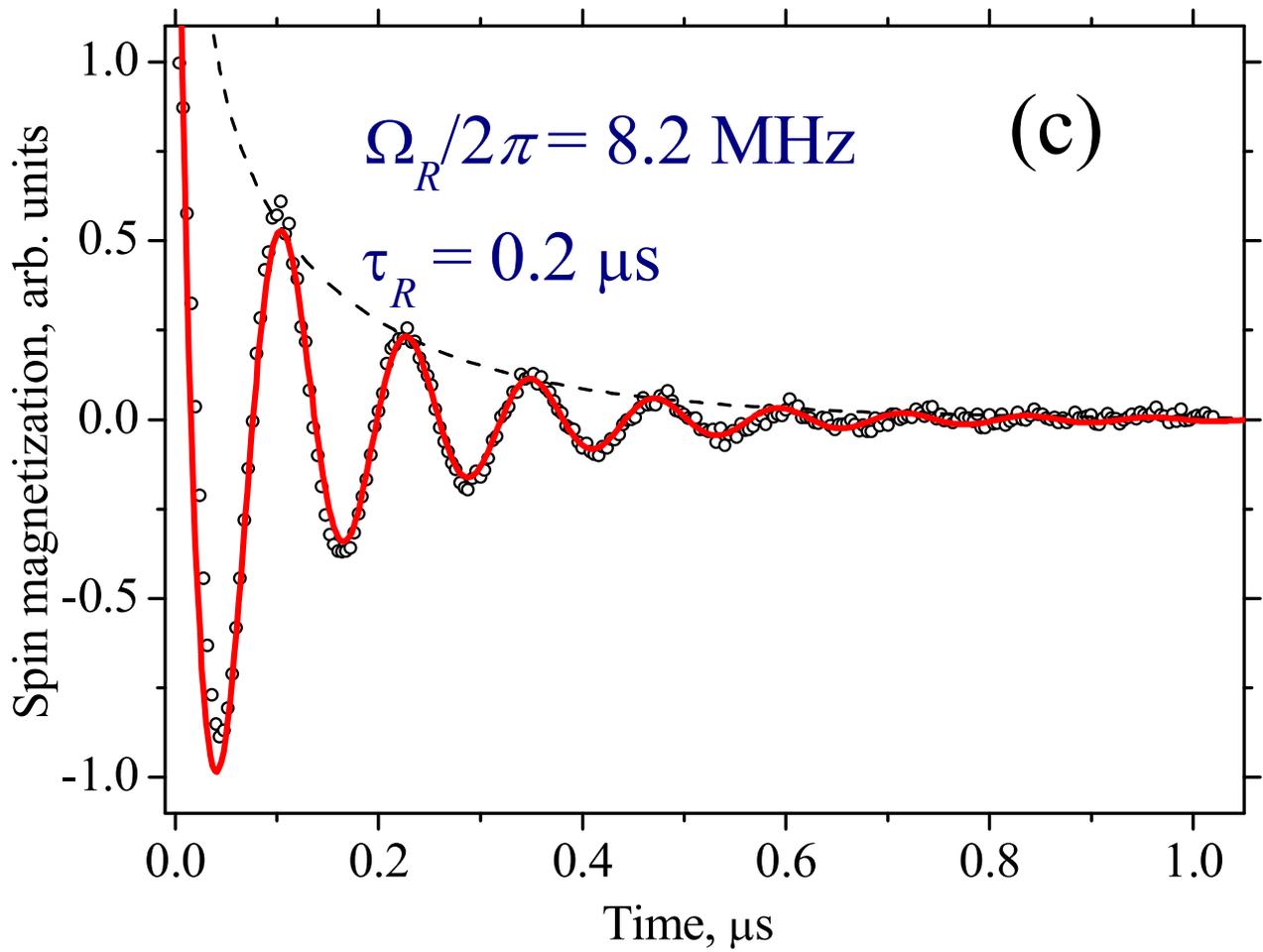

Figure 6c

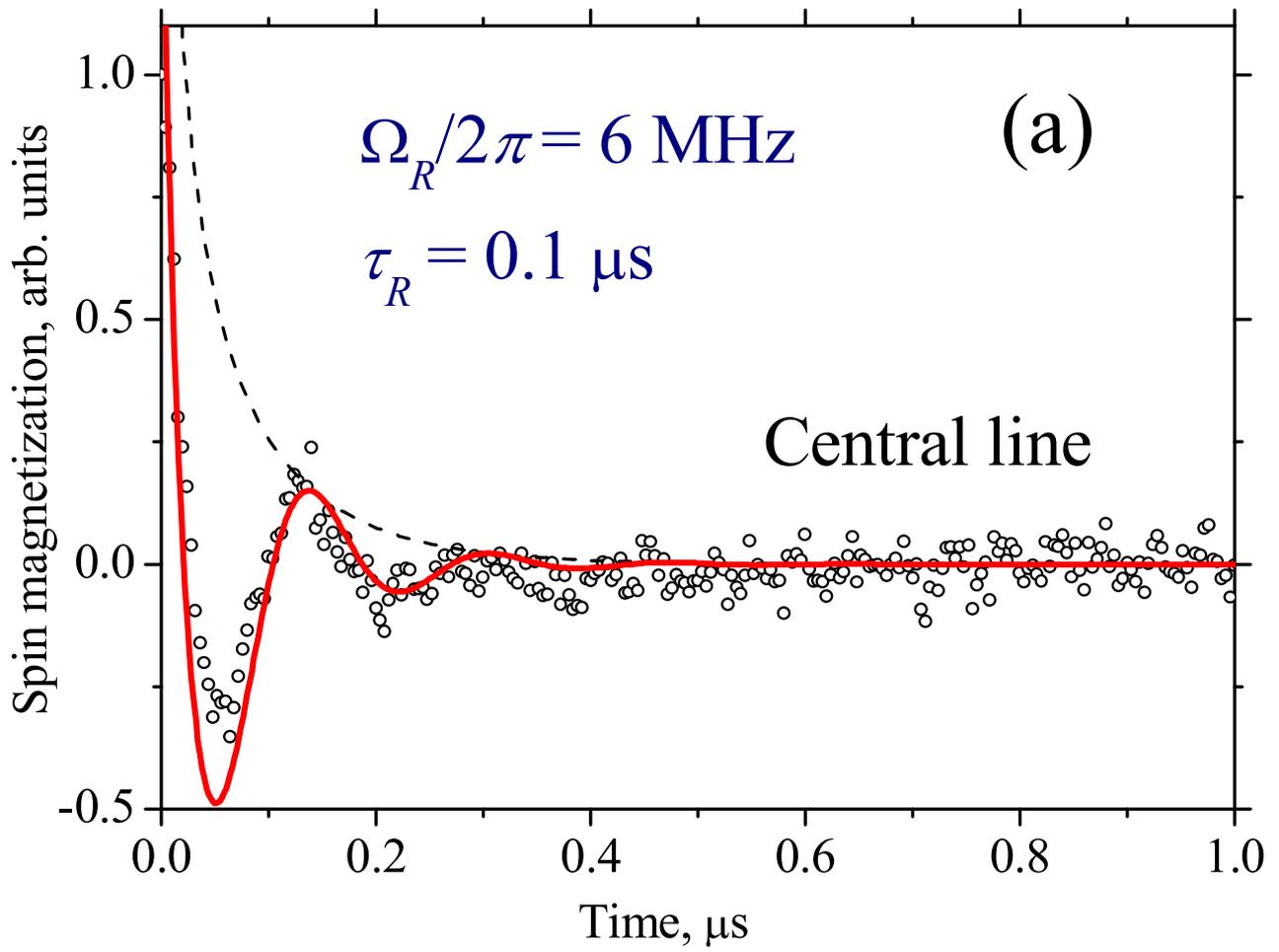

Figure 7a

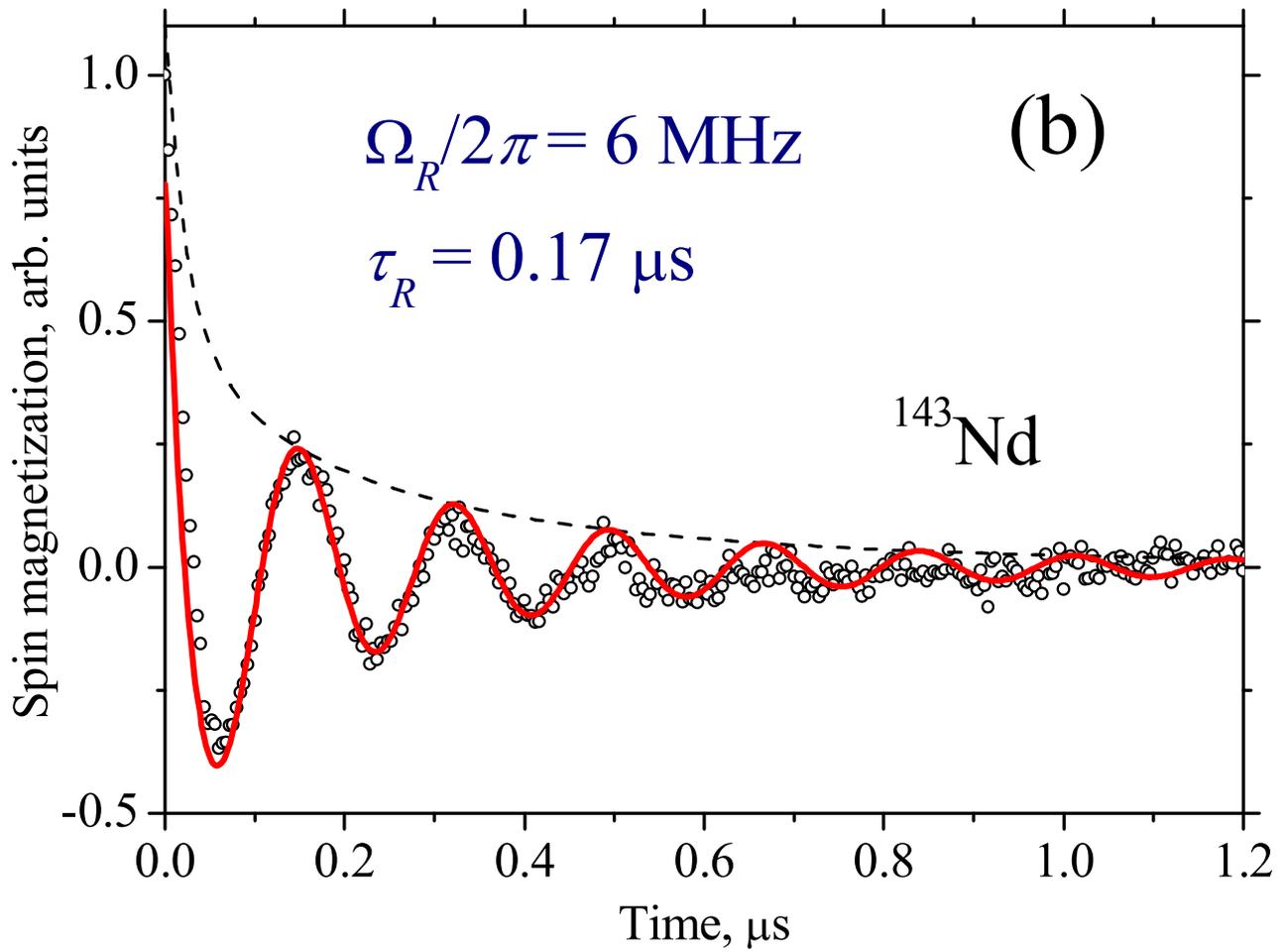

Figure 7b

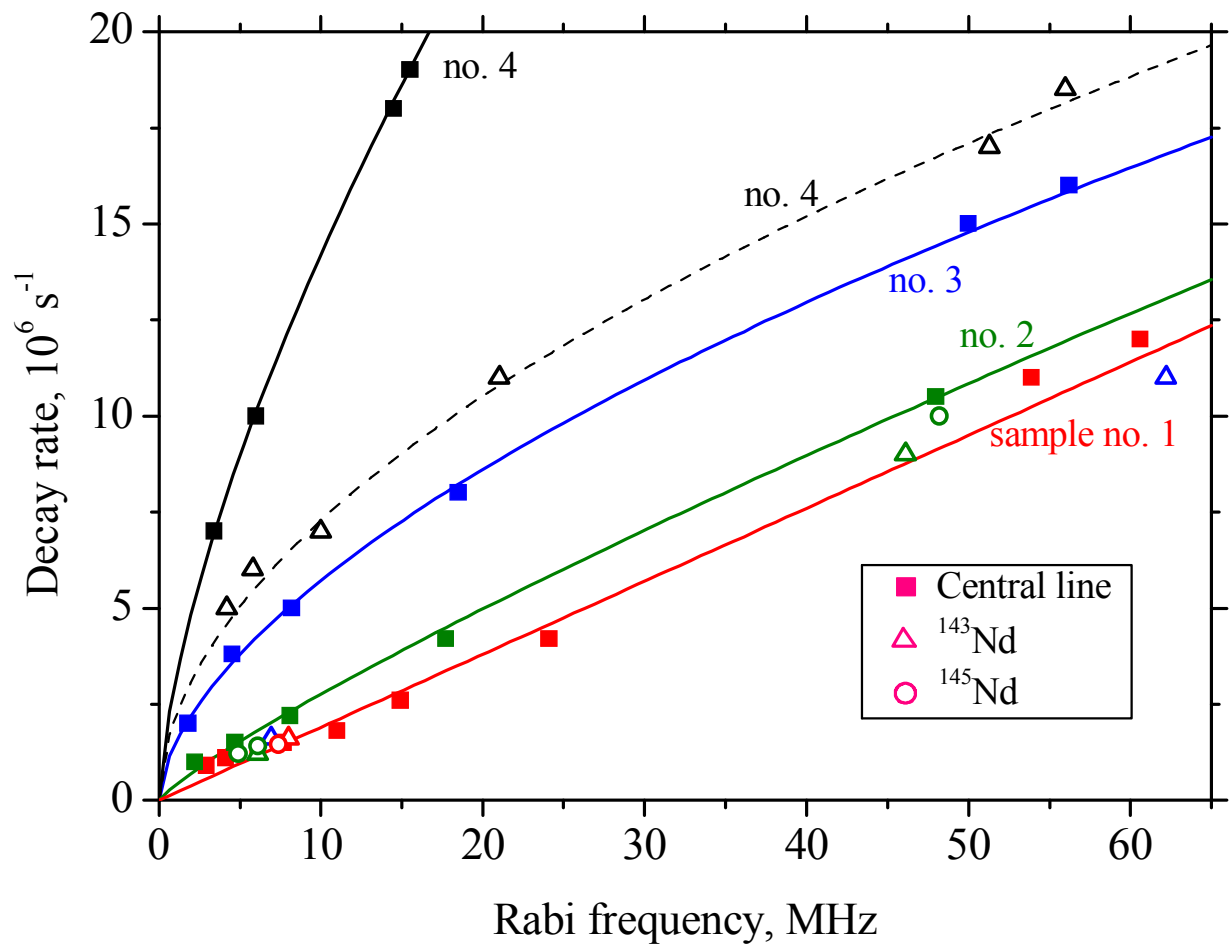

Figure 8

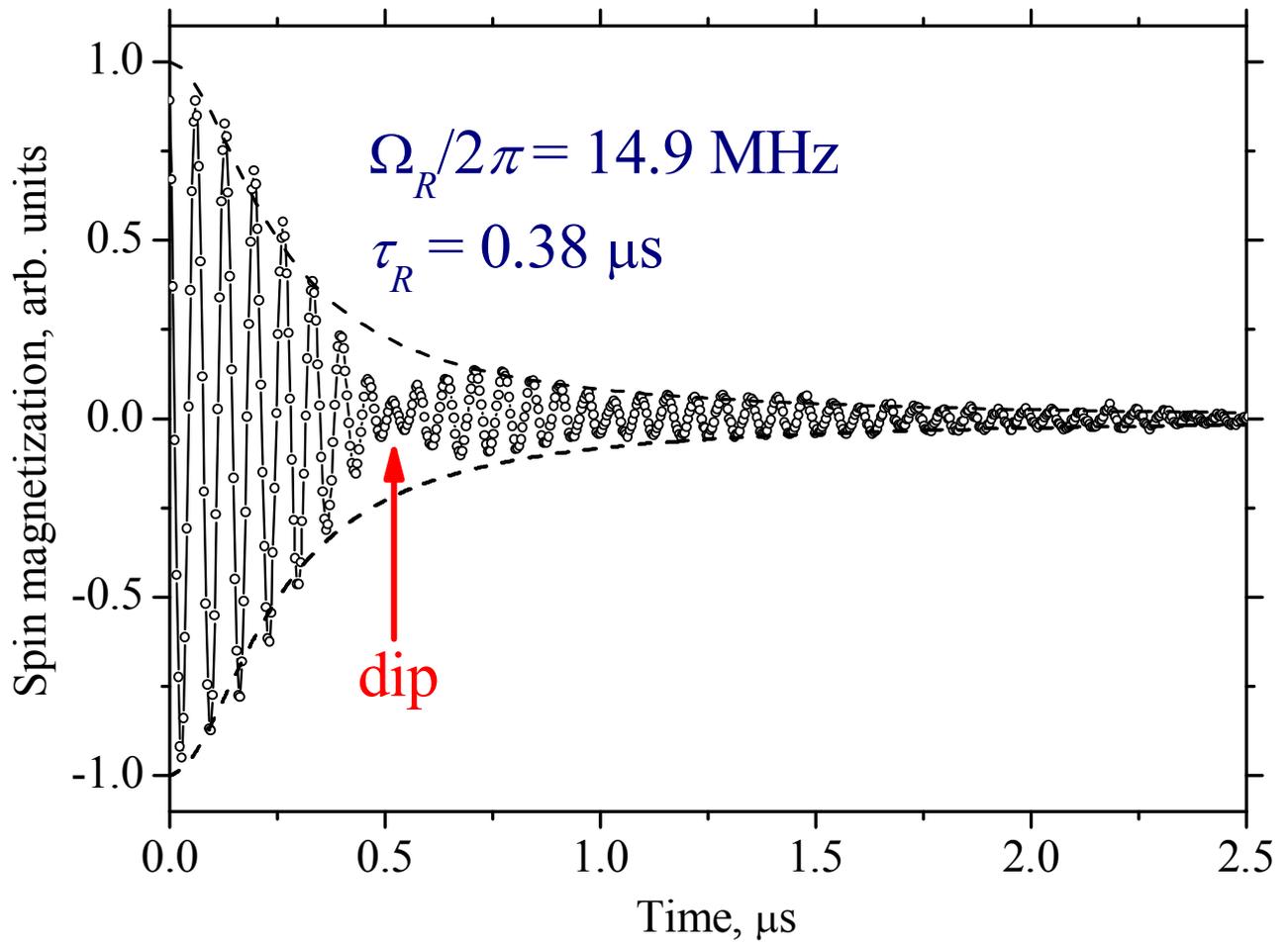

Figure 9